\documentclass[12pt]{iopart}
\usepackage{iopams}
\usepackage{charter}
\usepackage[latin1]{inputenc}   
\usepackage{graphicx}          
\graphicspath{{./}{figure/}}
\DeclareGraphicsExtensions{.eps,.png,.pdf}
\usepackage{dcolumn}           
\usepackage{nicefrac}          
\usepackage{bm}                
\usepackage[dvips]{color} 
\definecolor{red}{rgb}{0.85,0.1,0.0}
\definecolor{blue}{rgb}{0.0,0.0,1.0}
\definecolor{green}{rgb}{0.0,0.5,0.0}
\newcommand{\lfao}{La\-Fe\-As\-O}
\newcommand{\lfaoh}{La\-Fe\-As\-O$_{1-x}$\-H$_x$}
\newcommand{\lfaof}{La\-Fe\-As\-O$_{1-x}$\-F$_x$}

\newcommand{\tfmu}{TF-$\mu$SR}

\newcommand{\musr}{$\mu$SR}

\begin{document}
\title{Crossover between magnetism and superconductivity in low H-doped La\-Fe\-As\-O}
\author{G~Lamura$^1$, T~Shiroka$^{2,3}$, P~Bonf\`a$^4$, S~Sanna$^5$, R~De~Renzi$^4$, F~Caglieris$^1$, M~R~Cimberle$^1$,
S~Iimura$^6$, H~Hosono$^{6,7}$ and M~Putti$^1$}
\address{$^1$CNR-SPIN and Universit\`a di Genova, via Dodecaneso 33, I-16146 Genova, Italy}
\address{$^2$Laboratorium f\"ur Festk\"orperphysik, ETH-H\"onggerberg, CH-8093 Z\"urich, Switzerland}
\address{$^3$Paul Scherrer Institut, CH-5232 Villigen PSI, Switzerland}
\address{$^4$Dipartimento di Fisica e Scienze della Terra, Universit\`a  degli Studi di Parma, Viale G. Usberti 7A, I-43124 Parma, Italy}
\address{$^5$Dipartimento di Fisica and Unit\`a CNISM di Pavia, I-27100 Pavia, Italy}
\address{$^6$Materials and Structures Laboratory, Tokyo Institute of Technology, Yokohama 226-8503, Japan}
\address{$^7$Frontier Research Center, Tokyo Institute of Technology, Yokohama 226-8503, Japan}
\ead{tshiroka@phys.ethz.ch}
\date{\today}
\begin{abstract}
By a systematic study of the hydrogen-doped LaFeAsO system by means of dc resistivity, 
dc magnetometry, and muon-spin spectroscopy we addressed the question of universality 
of the phase diagram of rare-earth-1111 pnictides. In many respects, the behaviour of 
La\-Fe\-As\-O$_{1-x}$\-H$_x$ resembles that of its widely studied F-doped counterpart,
with H$^-$ realizing a similar (or better) electron-doping in the LaO planes. In a $x = 0.01$ 
sample we found a long-range SDW order with $T_{\mathrm{N}} = 119$\,K, while at 
$x = 0.05$ the SDW establishes only at 38\,K and, below $T_c = 10$\,K, it coexists at a
nanoscopic scale with bulk superconductivity.
Unlike the abrupt M-SC transition found in the parent La-1111 compound, the presence 
a \emph{crossover region} makes the H-doped system qualitatively similar to other 
Sm-, Ce-, or Nd-1111 families.
\end{abstract}

\submitto{\JPCM}
\maketitle

\section{\label{sec:intro}Introduction}
The complexity of high-temperature superconductors, reflecting their quantum-correlated nature,
has intrigued scientists over the years. In this respect the discovery of superconductivity below
$T_c = 26$~K in LaO$_{1-x}$F$_x$FeAs \cite{Kamihara2008} marked the beginning of a new era,
that of iron-based compounds, which could be tested against
the long-known cuprates. In both cases the doping of charges into the FeAs or CuO$_2$ layers,
respectively, plays a key role for the appearance of superconductivity.
The similar dependence of $T_c$ on carrier doping is reflected in the superconducting domes 
of Fe-based materials, qualitatively similar to those of many unconventional superconductors. 

Generally, in the phase diagrams of all the Ln1111 family members (Ln being a lanthanoid) 
the static magnetism of the parent compounds is suppressed by carrier doping (or pressure) 
in favour of the superconducting state.
Depending on the rare earth Ln, the way how this suppression effectively occurs gives
rise to different types of phase transitions. In the case of La, an abrupt transition from the magnetic (M) to the superconducting (SC) phase takes place \cite{Luetkens2009}.
On the other hand, an M-SC coexistence in the form of a mesoscopic phase separation was 
found in F-doped La-1111 samples with $x=0.06$ at ambient-\cite{Kadono2009} and with $x=0.055$ 
under hydrostatic pressure \cite{Khasanov2011}.
Similarly, for a whole series of other rare-earth metals (as, e.g., Ln=Ce \cite{Shiroka2011}, 
Sm \cite{SannaPRB2009}, Nd \cite{LamuraXXXX}) the transition is of a second-order type, with the M
and SC phases nanoscopically coexisting over a finite range of F doping. This picture changes
slightly in case of substitutions in the FeAs planes: here the nanoscopic coexistence of reentrant
magnetism with SC has been found in Ln1111 (Ln=La, Nd, Sm) \cite{Sanna2011,Sanna2013} and in
Ce1111 \cite{Prando2013} for Ru and Co substitutions, respectively. Nevertheless, if one
considers only substitutions in the LnO planes, it remains unclear what makes La so peculiar, i.e.,
if the strict separation of the M and SC phases is an intrinsic feature of this compound.

For a long time the low solubility limit of fluorine in 1111 systems ($x < 0.15-0.20$) has left
their over-doped SC region mostly unexplored. A clever way to circumvent this, at first sight
unsurmountable, problem was to make use of the high solubility of hydrogen \cite{Miyazawa2010}. 
Although counter-intuitive, neutron powder diffraction and DFT calculations have found
that hydrogen not only substitutes oxygen in the LnO layers but, most importantly, that
it adopts a $-1$ charge state. Hence, at all effects H$^-$ acts as if it were F$^-$, while
providing access to a very extended doping range [$0 < x < 0.5$ in (Ce, Sm)FeAsO$_{1-x}$H$_x$].
That hydrogen in its H$^-$ form substitutes the O$^{2-}$ 
and supplies electrons to the FeAs layers, exactly as fluorine does, is confirmed by the very
similar $T_c$ values found in the fluorine- and hydrogen-doped 1111 families in their common doping range.
In addition, in case of \lfaoh\ the extended H-doping range has revealed the presence of a
second superconducting dome peaked at \textit{x} = 0.3 with $T_c = 36$~K, not accessible
via F doping \cite{Iimura2012}. Very recently, both neutron diffraction and \musr\
measurements on the same family reported the discovery of a second antiferromagnetic phase,
 where iron spins form an antiferromagnetic collinear structure at very high doping level 
($x>0.4$). Interestingly, this stripe-type magnetic order coexists with superconductivity up to 
$x=0.45$ \cite{kadono2014}.

The availability of \lfaoh\ and a comparison with the well-known \lfaof\ family,
allowed us to address the initial question: is the abrupt M-SC transition in La-1111
intrinsic to La, or does it depend on other factors? To this aim we studied extensively two 
 H-doped samples \cite{Iimura2012} with low nominal hydrogen content $x(\mathrm{H})=0.01$ and 0.05 by
means of dc resistivity, magnetometry and muon-spin spectroscopy (\musr). In the 
$x(\mathrm{H})=0.01$ case we could detect a magnetically ordered phase below about 
$T_\mathrm{N}=119$\,K. The analysis of $\mu$SR data allowed us to establish a full equivalence 
between the hydrogen and fluorine doping, with our experimental results being in very good 
agreement with those of an  $x(\mathrm{F})=0.03$ sample \cite{JPCarlo2009}. In the $x(\mathrm{H})=0.05$ 
case, surprisingly, we found a short-range magnetically ordered phase, which extends over 
the whole sample volume and which coexists with bulk superconductivity. This unexpected 
result suggests that the presence of a M-SC crossover region could be a
common feature for the whole 1111 pnictide family.
\section{\label{sec:exp_details}Experimental details}
\subsection{\label{ssec:preparation}Sample preparation and characterization}
Two polycrystalline \lfaoh\ samples with nominal H content \textit{x}(H) = 0.01, 0.05 were synthesized by means of solid-state, high-pressure reaction, using La$_2$O$_3$, LaAs, LaH$_2$, FeAs and Fe$_2$As as starting materials, as reported in detail in \cite{Hanna2011,Iimura2012}. The sample \textit{x}(H) = 0.01 consisted of a single disk-shaped pellet (diameter 5.8 mm, average thickness 1.8 mm), whereas the sample \textit{x}(H) = 0.05 comprised a mosaic of pellets, whose biggest piece was a flat disk (diameter 5.7 mm, average thickness 1.3 mm). Room-temperature, powder x-ray diffraction measurements using Cu K$\alpha_1$ radiation were employed to assess the phase purity and to determine the structural parameters of the synthesized samples (see table~\ref{tab:impurities}). The main impurity consisted of unreacted La$_2$O$_3$, whose content never exceeded 2\% wt. To evaluate the exact hydrogen content in the synthesized samples we carried out thermal desorption spectroscopy measurements,
whereas
the chemical composition (excluding hydrogen) was determined by a wavelength-dispersive-type electron-probe microanalyzer.
\begin{table*}[tbh]
\centering
\caption{\label{tab:impurities}Lattice parameters and impurity content in \lfaoh\ as determined via x-ray diffraction.}
\begin{footnotesize}
\begin{indented}
\item[]\begin{tabular}{ccccc} 
\br
$x$(H) & $a$ (\AA) & $c$ (\AA) & Impurity type & (in wt \%) \\ \mr
0.01 &  4.034 &  8.733 &  La$_2$O$_3$     &  2 \\
0.05 &  4.031 &  8.725 &  La$_2$O$_3$     &  2 \\
\br
\end{tabular}
\end{indented}
\end{footnotesize}
\end{table*}
\subsection{\label{ssec:transport}Resistivity and magnetization measurements}
\begin{figure}[tbh]
\centering
\includegraphics[width=0.6\textwidth]{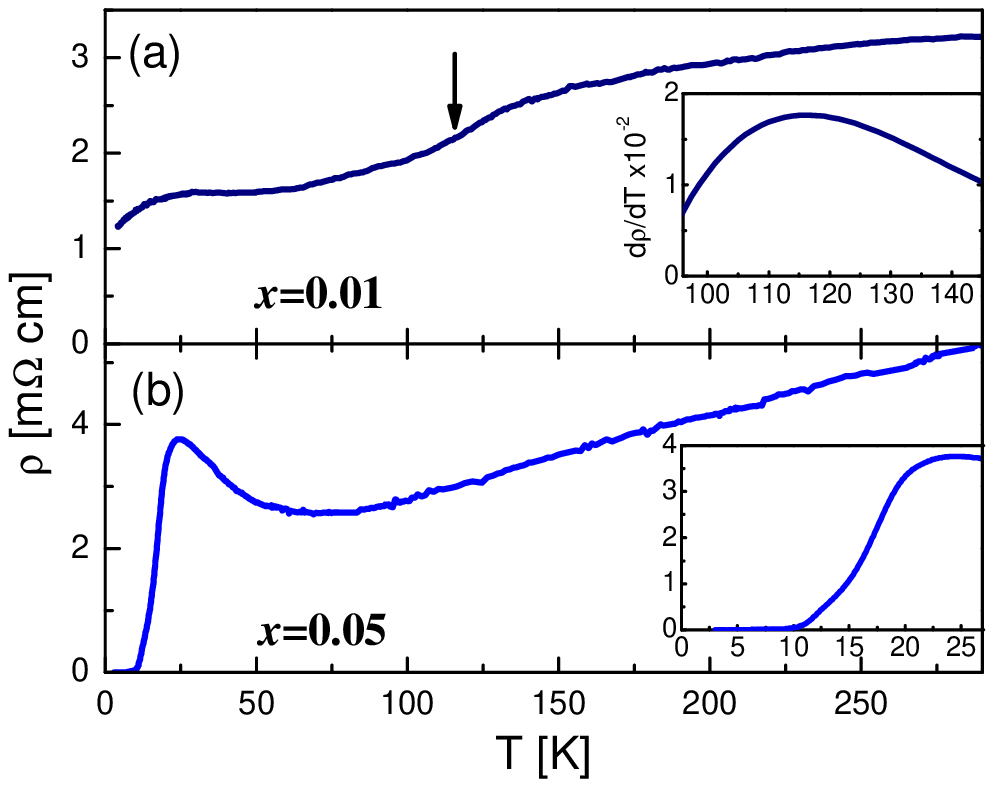} 
\caption{\label{fig:roALL}Resistivity vs.\ temperature in  \lfaoh\ for $x=0.01$ (a) and $x=0.05$ (b).
Insets show the maximum in $\mathrm{d}\varrho/\mathrm{d}T(T)$ (a) and the drop in $\varrho (T)$ near $T_c$ (b).}
\end{figure}

The resistivity of \lfaoh\ samples was measured by means of a standard four-point method, with
the temperature dependences $\rho(T)$ being shown in figure~\ref{fig:roALL}. Upon cooling, 
the sample $x(\mathrm{H})=0.01$  shows the typical transport features
of underdoped iron-based oxypnictides: a low-temperature resistivity in the
m$\Omega\,$cm range and an inflection point (arrow in figure~\ref{fig:roALL}),
defined as the maximum of the first derivative, $\mathrm{d}\rho/\mathrm{d}T$,
generally attributed to a spin-density wave (SDW) transition \cite{Hess2009}.
At very low temperatures ($T<20K$) we could detect also a small drop in resistivity. A similar 
feature has been found also in high-quality LaFeAsO single crystals, where
it was ascribed to a change of the magnetic structure of the iron atoms from an antiferromagnetic 
to a ferromagnetic arrangement along the $c$ axis \cite{Jesche2012}. The overall temperature behaviour
reproduces closely the results reported in \cite{Iimura2012} for a
nominally equivalent sample.
In the $x=0.05$ case the peak in $\mathrm{d}\rho/\mathrm{d}T$ disappears.
The resistivity decreases almost linearly down to 75 K, while below this temperature 
a weak localization takes place, just before the superconducting transition at ca.\ 10\,K, 
as defined by a zero-resistivity criterion.
Both these features are in good agreement with existing experimental data for 
an $x=0.04$ sample \cite{Iimura2012}.

\begin{figure*}[tbh]
\centering
\includegraphics[width=0.48\textwidth]{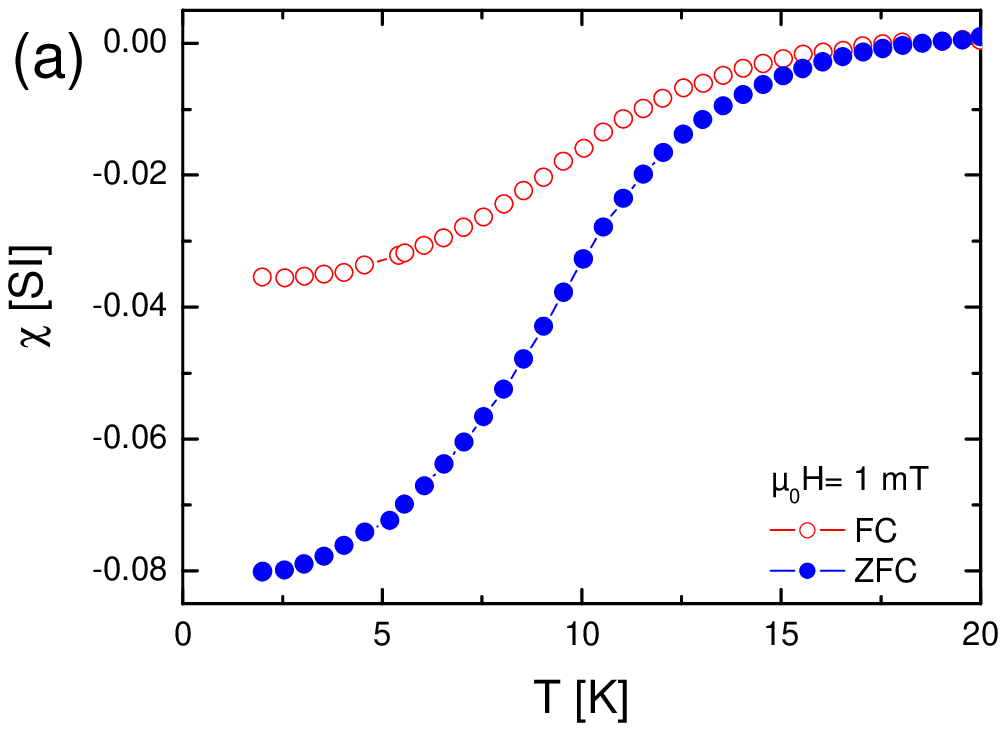} 
\includegraphics[width=0.46\textwidth]{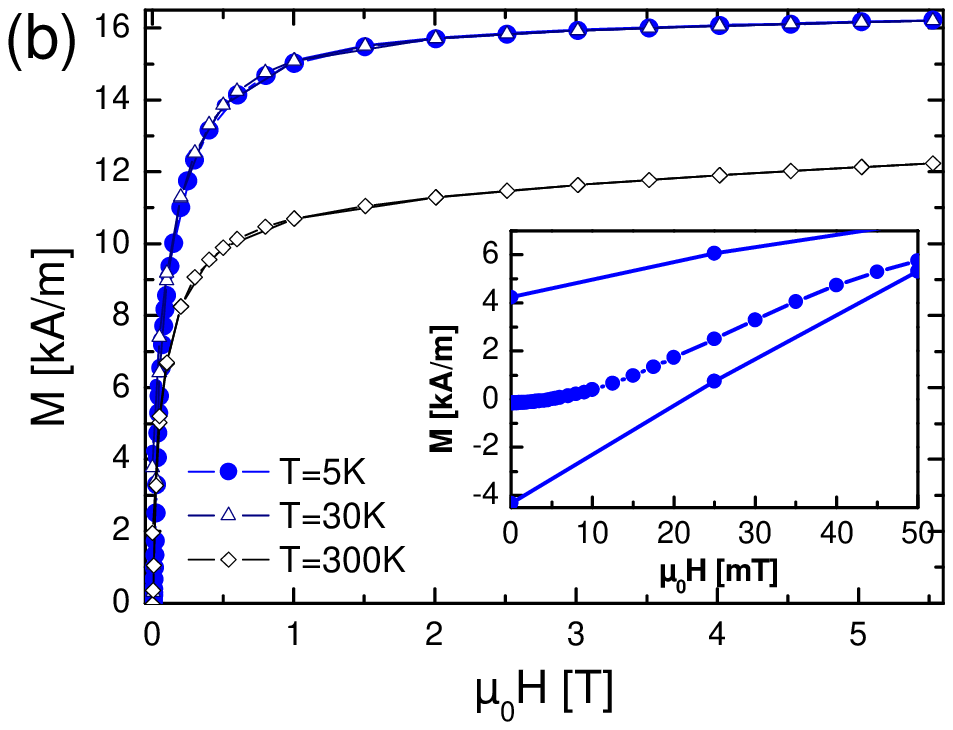} 
\caption{\label{fig:SQUID005}Magnetic characterization of the $x = 0.05$ sample: 
(a) zero-field cooled (ZFC) and field-cooled (FC) dc susceptibility at 1\,mT; 
(b) dc magnetization at 5, 30, and 300\,K. The inset shows the low-field region of the 5-K data. Both measurements were performed on a disk-shaped sample whose flat surface was parallel to the applied field. }
\end{figure*}

To characterize in more detail the superconducting state of the $x = 0.05$ sample, 
we carried out dc magnetization measurements by means of a superconducting 
quantum interference device (SQUID) magnetometer (Quantum Design).
The susceptibility vs.\ temperature curves were measured from 2 to 20\,K at 
$\mu_0 H = 1$\,mT, both in zero-field-cooled (ZFC) and in field-cooled (FC) conditions. 
In addition, dc isothermal magnetization measurements were performed at selected 
temperatures. The experimental results shown in figure~\ref{fig:SQUID005} can be 
summarized as follows:\\
\textit{i}) the shielding fraction as inferred from the ZFC susceptibility data 
shown in figure~\ref{fig:SQUID005}(a) (after subtracting a ferromagnetic offset) 
provides an apparent shielding fraction of about 8\%. 
This value, however, can be considered only as a lower bound for the 
superconducting volume fraction, since the attainment of the percolation threshold
(i.e., zero resistivity below $T_c$) indicates by itself that the superconducting phase
occupies at least 20--30\% of the sample volume \cite{Marck2009}.\\
\textit{ii}) the isothermal magnetization curves at 5, 30 and 300\,K indicate 
the presence of ferromagnetic impurities that mask
the intrinsic diamagnetic response of the sample. In particular, the first-magnetization
curve at 5\,K does not show a diamagnetic (i.e., superconducting) slope 
[see inset in figure~\ref{fig:SQUID005}(b)]. Instead, its initially constant behaviour results 
from the interplay of two opposite contributions, namely, a diamagnetic contribution 
from the superconducting phase and a ferromagnetic one from the above mentioned impurities.
\\
Besides the extrinsinc effects of diluted ferromagnetic impurities, these
experimental results resemble closely those regarding ruthenocuprates (see, e.g., 
\cite{Bernhard2000,Cimberle2003} and references therein), where superconductivity 
develops well within a magnetically ordered phase. Also in our case, we found a low-$T$ magnetic order in the FeAs planes (detected via $\mu$SR measurements --- 
see section~\ref{sec:discussion}), which coexists with a superconducting phase. 
Typical internal magnetic fields in the SDW state of pnictides are about
20--40\,mT \cite{SannaPRB2009,LamuraXXXX,Sanna2010}, i.e, significantly higher than 
$H_{c1}$, the first critical SC field ($\mu_0 H_{c1} \sim 6$\,mT at 1.8\,K in 
\lfaof\ polycrystalline samples) \cite{Kohama2009}. In these conditions, a 
spontaneous vortex phase can easily develop, hence preventing any meaningful 
ZFC measurements at temperatures below the onset of the magnetic order.\\
%
\subsection{\label{ssec:musr_exp}Muon-spin spectroscopy}
The muon-spin spectroscopy measurements were carried out at the GPS instrument of 
the S$\mu$S facility at the Paul Scherrer Institute, Switzerland. We performed both ZF, LF
and TF experiments. By means of the former one can detect spontaneous magnetic order,
as well as distinguish between short- and long-range order \cite{Shiroka2011,Yaouanc2011}.
The latter, instead, is typically used to distinguish between static and dynamic 
magnetism (on the $\mu$SR timescale) \cite{Drew2008,Yaouanc2011}. TF measurements were performed only on the \textit{x} = 0.05 sample to study its superconducting properties.

In all our experiments on both samples the relatively large sample thickness (about 1.5 mm) and the use of the veto counter, which ensures that the signal is acquired only from muons stopped in the sample, both contributed to significantly improve the signal-to-noise ratio. Below we summarize
the main experimental results for the $x = 0.01$ and 0.05 samples.
\begin{figure}[tbh]
\centering
\includegraphics[width=0.6\textwidth]{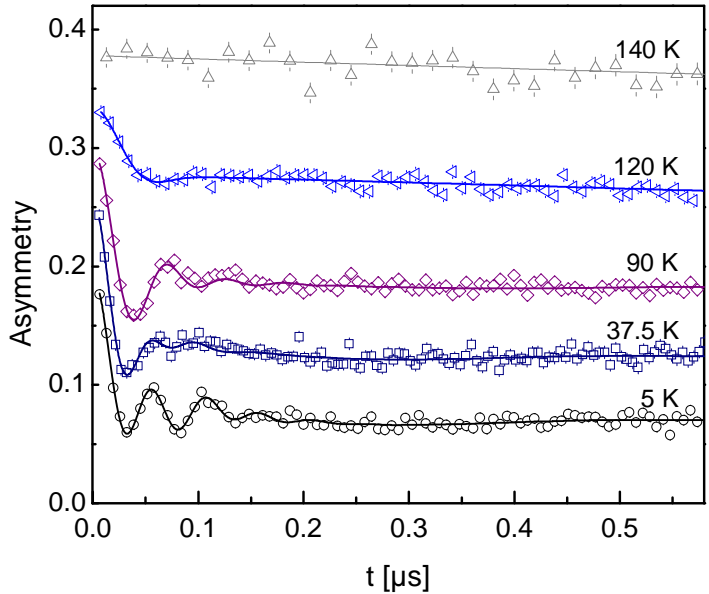} 
\caption{\label{fig:asymH001}ZF-$\mu$SR short-time spectra in \lfaoh\ for $x=0.01$ 
at selected temperatures. The highly damped oscillations observed at most temperatures 
are due to a superposition of different muon precession frequencies.} 
\end{figure}

\subsubsection{\label{ssec:h001}$\mu$SR in the $x(\mathrm{H})=0.01$ case:}
Figure~\ref{fig:asymH001} shows a selection of ZF-spectra for the $x = 0.01$ sample. 
The most prominent feature is the presence of highly damped oscillations, indicative of 
several different precession frequencies, reflecting the different local fields probed 
by the implanted muons. To distinguish these contributions we fitted the time-dependent 
$\mu$SR asymmetries with the following model:
%
%
\begin{equation}
\label{eq:spin_prec}
\frac{A^{\mathrm{ZF}}(t)}{A_{\mathrm{tot}}^{\mathrm{ZF}}(0)} =  \left[1-V_M(T)\right] \, g(t) +\sum_{i=1}^{N} w_i \cdot \left[a_{T_i} \, f_i(\gamma_{\mu}B_{\mu} t) \, D_{T_i}(t) + a_{L_i} \, D_{L_i}(t) \right],
\end{equation}
%
where $A_{\mathrm{tot}}^{\mathrm{ZF}}(0)$ is the high-temperature value of the initial asymmetry, 
$V_M$ is the magnetic volume fraction, $g(t)$ the time-dependent relaxation in the paramagnetic 
state, and $\gamma_{\mu} = 2\pi \times 135.53$\,MHz/T the muon gy\-ro\-mag\-ne\-tic ratio. 
In the magnetically ordered state a nonzero $V_M$ fraction of muons probes a local magnetic 
field $B_\mu$ at the implantation site $i$; $a_{T_i}$ and $a_{L_i}$ in equation~(\ref{eq:spin_prec}) refer to 
muons probing local magnetic fields in the transverse (T) or longitudinal (L) directions with 
respect to the initial spin polarization. 
The coherent precession of muons is taken into account by the $f(t)$ function, whereas 
$D_{T_i}(t)$ and $D_{L_i}(t)$ represent how this precession is damped. The former decay 
reflects the static distribution of local magnetic fields, whereas the latter is due to dynamical 
relaxation processes. Finally, the sum over \textit{i} generalizes equation~(\ref{eq:spin_prec}) to 
the case of several inequivalent crystallographic implantation sites, whose populations $w_i$ 
satisfy the normalization condition $\sum_i^{N} w_i =1$.

\begin{figure}[tbh]
\centering
\includegraphics[width=0.6\textwidth]{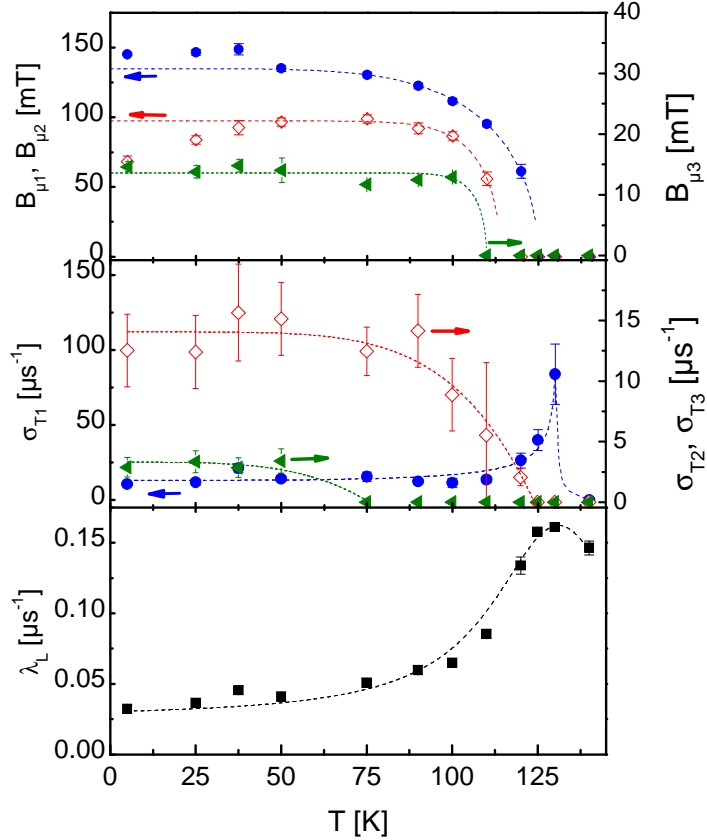} 
\caption{\label{fig:BmuH001}Temperature dependence of the internal magnetic fields $B_{\mu}^i$, their respective transverse Gaussian relaxation widths $\sigma_{\mu}^i$ and the longitudinal relaxation rates $\lambda_L$. The dashed lines are guides for the eyes.}
\end{figure}

In the specific case of $x = 0.01$ sample we find that (\textit{a}) the high-temperature
paramagnetic phase is best described by means of an exponential relaxation function 
$g(t)=e^{-\lambda t}$, which suggests the presence of fast fluctuating nuclear magnetic moments. 
(\textit{b}) In the magnetically ordered phase ($T <  T_\mathrm{N}$) three distinct precession
frequencies can reproduce the time-dependent asymmetry. The considerable damping of the time-dependent asymmetry signal cannot be fully taken into account by a heavily damped cosines function. For this reason the $f_i(t)$ (i=1,2,3) term was identified with a zeroth-order Bessel function. We recall that, the Bessel function is generally the fingerprint of incommensurate long-range magnetic order\cite{SAVICI,Yaouanc2011}.
The respective dampings, $D_{T_i}(t)$, were found to be of a Gaussian type. Finally, it was 
not possible to distinguish three different longitudinal relaxation rates. Therefore, 
these components were merged into a single, slowly-relaxing Lorentzian component. 
Figure~\ref{fig:BmuH001} and table~\ref{tab:properties} report all the parameters resulting 
from fits of the time-dependent $\mu$SR asymmetry by means of equation~(\ref{eq:spin_prec}).\\

\begin{table}[ht]
\caption{\label{tab:properties}Magnetic properties of the \lfaoh\ samples for $x=0.01$ and 0.05, as determined from $\mu$SR and dc magnetometry measurements. Field distribution widths were calculated as $\sigma_{T_i}/\gamma_{\mu}$ and $\lambda_{T_i}/\gamma_{\mu}$ for Gaussian and Lorentzian distributions, respectively. The values reported for $B_{\mu i}$ and $\Delta B_{\mu i}$ have been taken at 50 K and 5 K for \textit{x}=0.01 and \textit{x}=0.05 samples, respectively.}
\begin{footnotesize}
\begin{tabular}{cccccccccc} 
\br
$x$(H) & $T_c$ & $T_\mathrm{N}$\,(K) & $\Delta T_\mathrm{N}$\,(K) & $B_{\mu 1}$\,(mT) & $\Delta B_{\mu 1}$\,(mT) & $B_{\mu 2}$\,(mT) & $\Delta B_{\mu 2}$\,(mT) & $B_{\mu 3}$\,(mT) & $\Delta B_{\mu 3}$\,(mT) \\
\mr
0.01& -&119.3(7)& 6(1)&135(2)&17(2)&96.6(3)&18(4) &14.0(2)&4(1)\quad\\
0.05&10& 38.0(6)  &6.3(9)&--    &22(1)&--   &4.2(6)&--     &--    \quad\\
\br
\end{tabular}
\end{footnotesize}
\end{table}

\begin{figure}[tbh]
\centering
\includegraphics[width=0.6\textwidth]{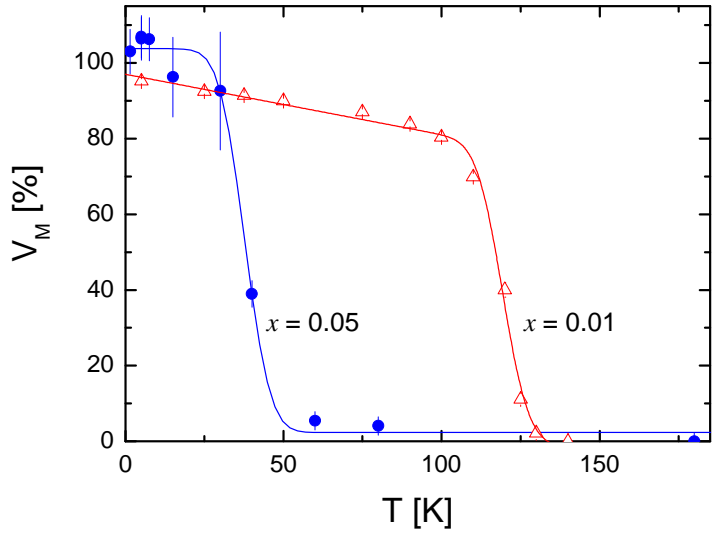} 
\caption{\label{fig:Vm_ALL}Temperature dependence of the magnetic volume 
fraction for the samples $x=0.01$ and 0.05. The continuous lines represent numerical fits 
using the \textit{erf} model function.}
\end{figure}

Figure~\ref{fig:Vm_ALL} instead shows the temperature dependence of the magnetic volume fraction. 
By fitting the latter with an error-function-like model, 
$1 - \mathrm{erf} \left[ (T - T_{\mathrm{N}} )/( \sqrt{2}\,T_{\mathrm{N}}) \right]$, 
superposed to a linear component (to account for a gradual saturation with decreasing temperature), one 
can deduce the average N\'eel temperature. 
In the $x = 0.01$ case, the magnetic volume fraction grows sharply up to 85\% at 100 K 
and then reaches gradually the 100\% value for $T<15$\,K. Such behaviour could be 
ascribed to the presence of small regions (less than 10--15\% of the sample volume) 
with slightly higher hydrogen content, as suggested also by resistivity measurement 
(see section~\ref{ssec:transport}). The results of the fits are summarized in table~\ref{tab:properties}.
\begin{figure}[tbh]
\centering
\includegraphics[width=0.6\textwidth]{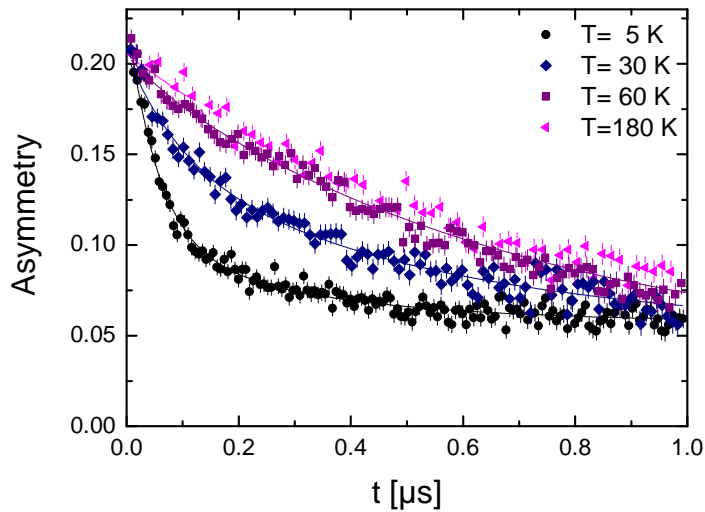} 
\caption{\label{fig:asymH005}ZF-$\mu$SR short-time spectra in \lfaoh\ for $x=0.05$ at selected temperatures. The significant increase of damping below ca.\ 60\,K suggests the onset of a magnetically ordered phase.}
\end{figure}
\subsubsection{\label{ssec:h005}$\mu$SR in the $x(\mathrm{H})=0.05$ case:}
The most representative ZF-spectra for the $x=0.05$ case are plotted in 
figure~\ref{fig:asymH005}. Two main features are evident: a rather large 
relaxation present also at high temperatures, and a significantly damped 
signal with no coherent precessions below about 60\,K. 
In this case, the time-dependent asymmetry was still fitted by means of  equation~(\ref{eq:spin_prec}), but with some differences with respect to the 
choices adopted for the sample $x = 0.01$:\\
\begin{figure}[tbh]
\centering
\includegraphics[width=0.65\textwidth]{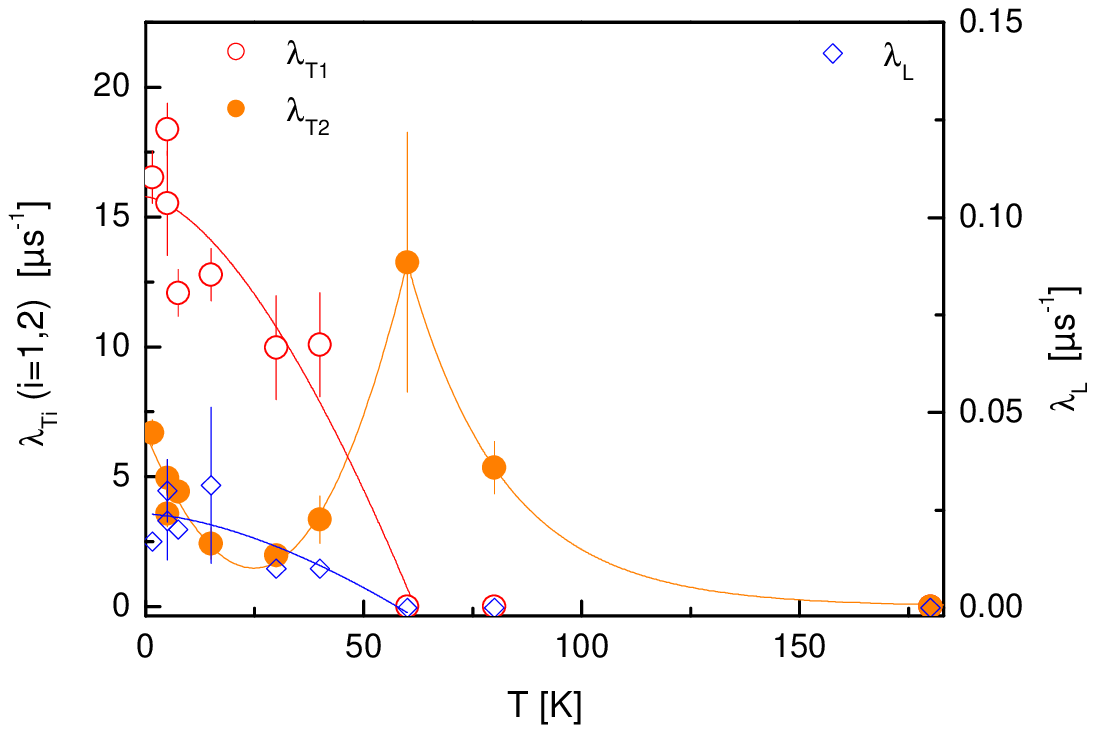} 
\caption{\label{fig:BmuH005}Temperature dependence of 
the transverse Lorentzian relaxation rates $\lambda_{T_i}$ ($i = 1$, 2) 
and the longitudinal relaxation rate $\lambda_L$ for the \textit{x}=0.05 sample. The continuous lines are guides for the eyes.}
\end{figure}
(\textit{i}) In the high-temperature paramagnetic phase, the $g(t)$ term is best 
described by a Lorentzian Kubo-Toyabe (KT) model, more suitable for fitting 
large relaxation rates due to homogeneously diluted ferromagnetic impurities (\textit{vide infra}).\\
(\textit{ii}) A very strong damping below ca.\ 60\,K indicates the onset 
of a magnetic order. Since the time-dependent asymmetry does
not oscillate, the product $f_i(t)D_{T_i}(t)$ was modelled with a 
decaying exponential.\\
Only two distinct transverse components could be detected, whereas it 
was not possible to distinguish those corresponding to the longitudinal 
relaxation. As before, such components were merged into a single 
slowly-decaying Lorentzian exponential. The parameters resulting from 
the fits using equation~(\ref{eq:spin_prec}) are summarized in 
figure~\ref{fig:BmuH005} and table~\ref{tab:properties}.
Figure~\ref{fig:Vm_ALL} reports again the temperature dependence 
of the magnetic volume fraction. Also in this case, the average magnetic 
ordering temperature was deduced by a fit with an error-function-like 
model (see table~\ref{tab:properties}). Differently from the $x = 0.01$ 
case, the $x = 0.05$ sample shows a very broad magnetic transition and 
seems to fully order only at the lowest temperature.  Moreover, no additional 
linear terms were necessary for the fit.

\section{\label{sec:discussion}Discussion}
\subsection{\label{ssec:doping}Low vs.\ intermediate H doping}
\textit{\textit{x}=0.01 sample: }
Normally, the signature of a magnetically ordered phase in ZF-$\mu$SR 
experiments is an oscillating signal representing the fingerprint of a coherent 
muon-spin precession around the local field at the implantation site. This is the 
case for all the parent compounds ($x = 0$) of the Ln-1111 iron pnictides, 
as well as for our lightly-doped $x = 0.01$ sample (see figure~\ref{fig:BmuH001}).
As in the case of Ce-1111 family \cite{Shiroka2011}, even tiny amount of doping contents are sufficient for changing the commensurate AF order of the parent compound into an incommensurate one, as suggested by the Bessel best fit of the time dependent-asymmetry \cite{SAVICI,Yaouanc2011}. Differently from the parent compounds, where two distinct frequencies,
reflecting muons implanted in the FeAs and LaO planes, were predicted \cite{Maeter2009} 
and found experimentally \cite{Maeter2009,Luetkens2009}, in our case 
we detect three different precessions at ca.\ 145, 68, and 15\,mT (corresponding to 18.3, 13.1 and 1.9 MHz) with relative weights $w_1$, $w_2$, and $w_3$ of 0.69, 0.20, and 0.11, respectively (see 
table ~\ref{tab:properties}).
As generally accepted for the Ln-1111 family \cite{Maeter2009}, muons implanted 
in the FeAs planes represent the most populated site. By the same token, our 
$B_{\mu 1}$ and $B_{\mu 3}$ fields are most likely attributed to muons stopping in FeAs and LaO planes, respectively.
These assumption are in very good agreement with results from a 3\% fluorine-doped
\lfao\ sample \cite{JPCarlo2009}, indicating that 3\%-F and 1\%-H substitutions 
correspond to similar effective doping levels. This analogy is further reinforced 
by considering that the N\'eel temperature of the $x(\mathrm{H}) = 0.01$ sample, 
as resulting from fits of the $T$-dependence of the magnetic volume fraction, 
fully agrees with the one of the fluorine-doped sample \cite{JPCarlo2009}.
At the same time, the presence of third frequency ($B_{\mu 2}$), surviving down to almost T$_N$, is somehow unexpected. In fact, contrary to theoretical predictions \cite{Maeter2009}, this seems to imply a third muon implantation site. Most likely, local-doping inhomogeneities might reduce the probed field at the standard muon implantation sites and significantly broaden magnetic transitions, as confirmed by a ~10-15\% fraction of sample volume whose magnetization never saturates below T$_N$, giving rise to a rather unusual linear increase in the magnetic volume fraction with decreasing temperature (see figure ~\ref{fig:Vm_ALL}).
In conclusion, given the similar behaviour of the $x (\mathrm{H}) =0.01$ and 
$x(\mathrm{F}) = 0.03$  systems (the latter being a nominal content), it seems that the 
O-H substitution dopes the Ln-1111 system more effectively with electrons.

\textit{\textit{x}=0.05 sample:}
At intermediate dopings a short-range magnetic order, evidenced by a  
fast-decaying non-oscillating asymmetry signal, is usually found either in 
F-doped \cite{SannaPRB2009,Shiroka2011}, or in Fe-substituted samples 
(e.g., Fe-Ru substitution \cite{Bonfa2012,Sanna2013}). 
This seems to be the case also for the $x = 0.05$ sample, whose $\mu$SR 
relaxation data reported in figure~\ref{fig:asymH005} belong to two distinct 
temperature regimes: below and above $T_{\mathrm{N}}$, respectively.
In the low-$T$ regime, the short-range magnetism implies a fast dephasing 
of muon spins, resulting in a sharp drop of the asymmetry. The large dephasing 
is due to local fields with widths $\Delta B_{\mu} = \lambda_{T_i} / \gamma_{\mu}$ 
of ca.\ 22 and 4 mT at the lowest temperature. Since the relative weights 
are $w_1= 0.74$ and $w_2 = 0.24$, their ratio recalls those generally accepted 
for the  muon populations implanted in the FeAs and LnO planes of 1111 family.

To verify the static nature (on the $\mu$SR timescale) of the local fields we 
performed a longitudinal-field (LF) spin-decoupling experiment at 5\,K.
\begin{figure}[ht]
\centering
\includegraphics[width=0.6\textwidth]{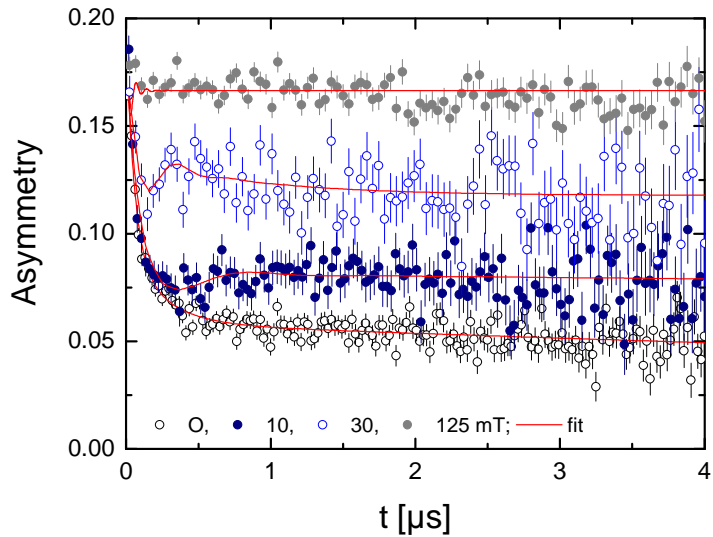}       
\caption{\label{fig:H005_LF5K}Longitudinal-field $\mu$SR asymmetry data at $T=5$\,K 
for the $x=0.05$ sample. Solid lines represent fits using the particular model adopted 
for $T<T_{\mathrm{N}}$ (see text for details).}
\end{figure}
An external magnetic field of increasing magnitude $B_{\parallel}$ is applied 
along the initial muon-spin direction. As long as $B_{\parallel}$ is lower than 
or of the same order of the internal static fields, it has a negligible influence 
on the muon polarization. However, when $B_{\parallel}$ exceeds the typical 
internal field value, it ``quenches'' the muon spins along the field direction 
and determines the full recovery of the longitudinal fraction. This is not the 
case, for strongly fluctuating internal fields, where the effect of the external field 
is barely noticeable. Figure~\ref{fig:H005_LF5K} shows 
the time-dependent asymmetry at different applied longitudinal fields.
Clearly, an applied field above ca.\ 50 mT is sufficient to fully recover the total 
muon-spin polarization. This is confirmed by the results shown in 
figure~\ref{fig:H005_LF5K_AMP}, where the initial longitudinal asymmetry $a_L$ 
and the transverse relaxation rates $\lambda_T$ are plotted as a function of $B_{\parallel}$. 
The saturation of the former and the disappearance of the latter unambiguously 
confirm the static nature of the internal fields in the low-$T$, magnetically-ordered 
phase of the $x = 0.05$ sample.
\begin{figure}[ht]
\centering
\includegraphics[width=0.6\textwidth]{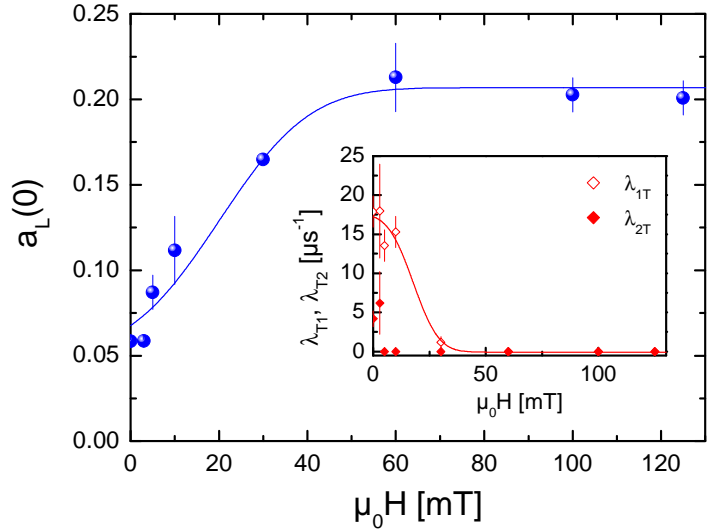}       
\caption{\label{fig:H005_LF5K_AMP}Initial longitudinal fraction vs.\ applied 
field at $T=5$\,K for the $x=0.05$ sample. Inset: field dependence of the transverse 
relaxation rates.  The continuous lines are guides for the eyes.}
\end{figure}

\begin{figure}[tb]
\centering
\includegraphics[width=0.6\textwidth]{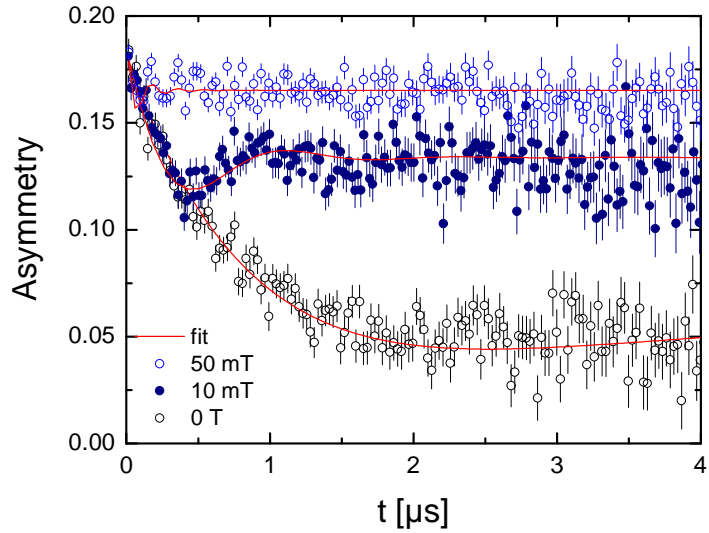}
\caption{\label{fig:H005_LF240K}LF-$\mu$SR time-domain spectra measured 
at $T = 240$~K in the $x=0.05$ sample. The continuous lines represent numerical fits 
with a Lorentzian Kubo-Toyabe model.}
\end{figure}

Above $T_{\mathrm{N}}$ the asymmetry behaviour does not depend on temperature 
and its time evolution is well described by a Lorentzian Kubo-Toyabe model, generally
indicative of the presence of diluted ferromagnetic impurities \cite{Sanna2009}. 
To confirm this hypothesis, another series of LF decoupling experiments was 
carried out at $T = 240$\,K.
Figure~\ref{fig:H005_LF240K} shows the time-dependent asymmetry at different 
applied fields, with all the data sets fitted using the above-mentioned field-dependent 
model. The prompt recovery of the full asymmetry in a relatively small applied field (50 mT) 
confirms the above picture. A posteriori, this result justifies also the macroscopic 
magnetization behaviour as evidenced by dc-magnetization measurements 
(see section~\ref{ssec:transport}).

\subsection{\label{ssec:pinning}Flux pinning and superconducting volume fraction}
Transport and dc susceptibility measurements show that the $x=0.05$ sample 
is a superconductor with $T_c = 10$\,K (see table~\ref{tab:properties}) 
and a superconducting volume fraction $V_{SC}$ of at least 20--30\%.
However, the estimate of the true $V_{SC}$ from macroscopic data is notoriously
difficult in complex systems having a magnetically ordered phase \cite{Bernhard2000} with
$T_{\mathrm{N}} > T_c$ and, occasionally, also diluted magnetic
impurities. In our case, things are even more complicated, since $H_{c1}$ is lower than 
the internal magnetic field due both to the ordered Fe$^{2+}$ moments in the 
SDW phase and to the remanent magnetization from the
diluted ferromagnetic impurities.

In such a situation even a conventional microscopic approach would still be 
faced with difficulties. Nevertheless, by the use of the so-called ``\tfmu\ 
pinning technique'' \cite{Bernhard2012}, we could overcome the difficulties 
related to the $V_{SC}$ determination, and find truly bulk superconductivity in
the $x = 0.05$ sample. \tfmu\ measurements were carried out by
applying a magnetic field perpendicular to the muon momentum $\boldsymbol{p}_\mu$.
In this geometry, the \textit{flux pinning} experiment consisted in
field-cooling the sample in an external field $B_{\mathrm{ext}} = 100$\,mT 
then, while keeping the temperature fixed at 5\,K (i.e., below $T_c$), in 
subsequently increasing the field by $\sim 25\%$ to 125\,mT. This approach 
allowed us to study the SC vortex lattice by observing the changes in 
the respective $\mu$SR spectra.

\begin{figure}[tb]
\centering
\includegraphics[width=0.7\textwidth]{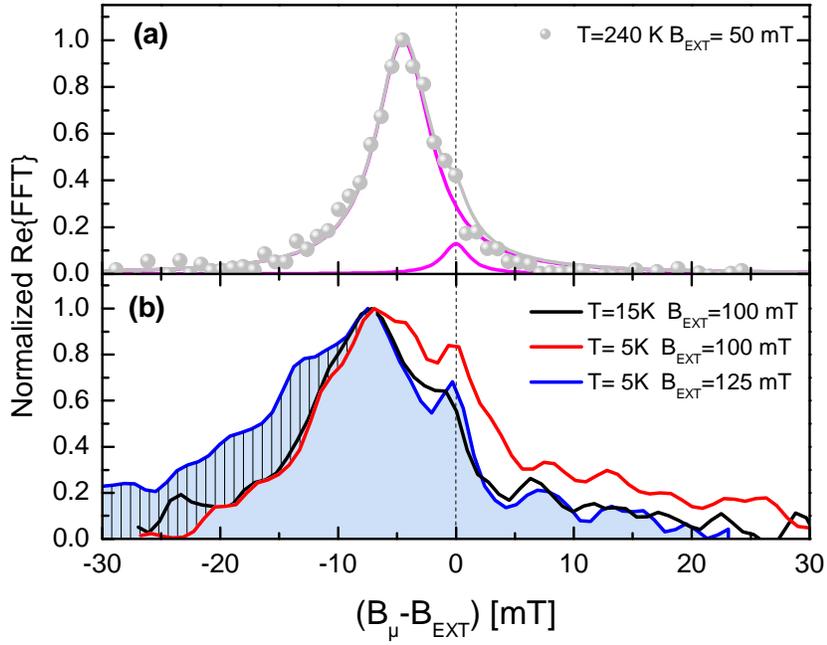}
\caption{\label{fig:normFFT}Normalized real-FFT amplitude vs.\ ($B_{\mu}-B_{\mathrm{ext}}$) 
for four \tfmu\ measurements at different temperatures and applied fields. 
(a) TF data at $T = 240$\,K and $B_{\mathrm{ext}}=50$\,mT (full circles). Continuous magenta 
lines represent two distinct Lorentzian fit components, with the grey curve being their sum. 
(b) TF data at 15\,K and 100\,mT (black line), at 5\,K and 100\,mT (red line), and at 5\,K and 125\,mT (blue line). 
The hatched region represents the effects of pinning on the field distribution of the SC vortex lattice  
once the applied field is raised to 125 mT (with $T = 5$\,K being kept fixed). The vertical dashed line 
indicates the unshifted normal-state component. See text for details.}
\end{figure}

To understand the outcome of the pinning experiment, it is useful to consider 
first what happens well above $T_{\mathrm{N}}$. 
Figure~\ref{fig:normFFT}(a) displays the fast Fourier transform (FFT) of the 
muon asymmetry data at 240\,K in a 50\,mT TF configuration. The amplitude of 
the real-FFT signal is proportional to the local-field distribution $P(B)$, 
which in our case is characterized by a main Lorentzian peak, shifted by about 
$-4.3$\,mT, and a small shoulder, centered at the applied-field value.
This picture is confirmed by a fit of the time-dependent asymmetry with the 
model:
\begin{equation}
\label{eq:TF_spin_prec}
\frac{A^{\mathrm{TF}}(t)}{A_{\mathrm{tot}}^{\mathrm{TF}}(0)} =  \sum_{i=1}^{N} a_i \, \cos(\gamma_{\mu}B_{\mu}^i t) \, \exp{(-\lambda_i t)}.
\end{equation}
Only two components ($N = 2$) were required for the fit, of which one at zero 
shift ($B_{\mu} - B_{\mathrm{ext}} = 0$) and with a relative weight of 3\%, in 
good agreement with x-ray diffraction results, that indicate the presence of 
2\% of non-magnetic La$_2$O$_3$ (see table~\ref{tab:impurities}). The absence 
of other fit components not only confirms that the sample is in a single phase, 
but also that the ferromagnetic impurities can be considered as homogeneously 
diluted nanoscopic inclusions, whose amount is below the detection threshold 
of x-ray diffraction. The unexpected diamagnetic shift of the main component 
[see figure~\ref{fig:normFFT}(a)], can be accounted for by considering the local 
magnetic field as seen by the implanted muons:
%
\begin{equation}
\label{eq:loc_field}
B_{\mu}=B_{\mathrm{ext}}+B_{\mathrm{demag}}+B_{\mathrm{L}}+B_{\mathrm{dip}}+B_{\mathrm{hf}}.
\end{equation}
The ferromagnetic impurities can be considered as homogeneously diluted magnetic 
moments mostly aligned along $B_{\mathrm{ext}}$. In this highly symmetric condition 
the sum of the dipolar contributions $B_{\mathrm{dip}}$ from the impurities
within the sphere of Lorentz construction  should be negligible. Moreover,
in a first instance, we neglect the hyperfine field $B_{\mathrm{hf}}$, as well 
(mostly because of the significant distance between an implanted muon
and an impurity).
We can, hence, focus on the two non-negligible terms: the demagnetizing field 
$B_{\mathrm{demag}}=-N_{\parallel}\,\mu_0 M_{\parallel}$~\footnote{In
\tfmu\ the applied field was orthogonal to the sample surface, i.e., parallel to the disk axis of symmetry. For this reason, by following the literature \cite{Pool,Cardwell,Beleggia2006}, we refer to the demagnetizing factor as $N_{\parallel}$.} and $B_{\mathrm{L}}=(1/3)\,\mu_0 M_{\parallel}$,
the Lorentz counter-sphere contribution. The magnetization $M_{\parallel}$ was measured at 240 K 
by dc magnetometry on the same disk-shaped sample\footnote{Diameter and thickness 
equal to 5.68 mm and 1.26 mm, respectively. This was the main piece which composed the 
mosaic sample used for the \musr\ measurements.} of figure~\ref{fig:SQUID005}, with the 
applied field in the same geometry of the \tfmu\ experiment.
For this configuration the demagnetization factor was estimated to be 
$N_{\parallel}\approx0.9$.\footnote{Firstly we estimated $N_{\parallel}\approx0.79$ by
approximating the sample to an oblate spheroid \cite{Pool,Cardwell}. Yet, this value 
can be as high as 0.85, since the oblate spheroid represents only a lower-bound 
approximation for a disk-shaped sample \cite{Beleggia2006}. Since the sample itself, 
in any case, is not a regular geometric body, the ultimate $N_{\mathrm{demag}}$ value 
can only be determined experimentally. This was done by measuring the sample's magnetic 
response in a parallel- and in a perpendicular-field geometry. Since the values 
$N_{\parallel}\approx0.9$ and $N_{\perp}\approx0.05$ provide the same intrinsic 
susceptibility they were chosen as our best estimates for $N_{\mathrm{demag}}$.}
Therefore, from the experimental value $\mu_0 M_{\parallel}(240\,\mathrm{K}, 50\,\mathrm{mT}) = 7.3$\ mT,
we could estimate a lower bound for the expected shift $B_{\mu} - B_{\mathrm{ext}} = B_{\mathrm{demag}} + B_{\mathrm{L}} \approx -4.2$\,mT, in excellent agreement with -4.3\,mT, the diamagnetic shift value
obtained from FFT data (see figure~\ref{fig:normFFT}(a) and table~\ref{tab:shift}).\\
\begin{figure}[tb]
\centering
\includegraphics[width=0.6\textwidth]{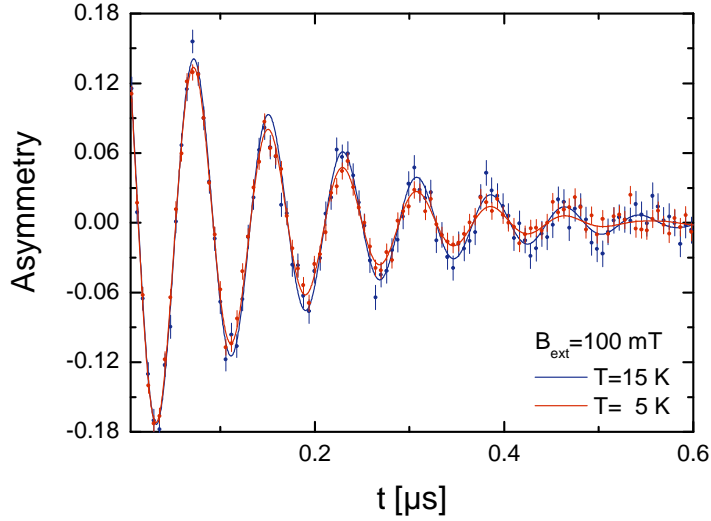}
\caption{\label{fig:TFasym}Short-time TF asymmetry data at 15 and 5 K under $B_{\mathrm{ext}}=100$\,mT.
The continuous lines represent two-component fits by means of equation~(\ref{eq:TF_spin_prec}).}
\end{figure}
We can now consider the experimental results of the \textit{flux-pinning}
experiment. Table~\ref{tab:shift} reports the diamagnetic shift expected at low temperature, 
highlighting the good agreement between the measured and calculated values. 
Figure~\ref{fig:normFFT}(b) shows the initial local-field distribution $P(B)$
at $T = 15$\,K and $B_{\mathrm{ext}}=100$\,mT (black line).\footnote{In a standard \emph{flux-pinning} experiment one usually plots $P(B)$ vs.\ $B_\mu$, a choice
which leaves \emph{unchanged} the field distribution profile in case of small
increases of the applied field (see, e.g., figure~5 in \cite{Bernhard2012}).
In our case, by choosing $B_{\mu} - B_{\mathrm{ext}}$ as the $x$-axis, the field profile
shows an \emph{apparent} left shift when the field is increased [hatched area in figure~\ref{fig:normFFT}(b)]. } The position
of the secondary component (identified with La$_2$O$_3$) remains unchanged
with respect to the high-temperature data (i.e., it follows closely the applied field), 
whereas most of the sample lags behind and shows a diamagnetic shift with respect to the 
high-$T$ data, reflecting an increase in bulk magnetization (see table~\ref{tab:shift}). 
Since at an applied field $B_{\mathrm{ext}}=100$\,mT the AF ordering is fully quenched 
(see figure~\ref{fig:H005_LF5K_AMP}), no other significant effects are expected.

Once the sample is field-cooled below $T_\mathrm{c}$, to 5\,K, the width of the main component (red line) increases mostly on the right side of the main peak as expected in case of a vortex lattice field modulation that generally induces an asymmetric modification of the local field profile \cite{Yaouanc2011}. Indeed, at low temperatures the depolarization rate grows significantly up to 6.5\,$\mu s^{-1}$ (see figure~\ref{fig:TFasym} and table~\ref{tab:shift}). These effects reflect the build up of a flux-line lattice in the SC state that further modulates the local field at the muon site.

If the SC volume fraction were as low as the minimal value
required by the percolation-threshold criterion, we would expect two possible scenarios: (\textit{i}) a third precessing component with a decay rate lower than $\sim 1$\,$\mu s^{-1}$, as reported for under-doped \lfaof\ superconducting samples \cite{Luetkens2008PRL}; (\textit{ii}) no variations in the local field profile \textit{P(B)}, because the considerable depolarization rate measured at 15 K (5.5\,$\mu s^{-1}$) could, in principle, fully mask other slow-decaying components.
To our surprise, fortunately neither case was verified: we measured a sizeable increase of the transverse 
depolarization rate and the fit of the time-dependent asymmetry data at 5\,K 
still required only two components (see figure~\ref{fig:TFasym}), one of which 
with a relative weight of 2.4\%, previously ascribed to a non-magnetic impurity phase.
Therefore, since the contribution from the diluted ferromagnetic impurities is constant at
a constant applied field, an increase in depolarization rate of the main component below
$T_c$ can only be associated to a superconducting phase whose vortex lattice affects the \emph{entire sample volume}.

\begin{table*}[tbh]
\centering
\caption{\label{tab:shift}Summary of the main fit parameters as resulting from fits of TF-$\mu$SR data with equation~(\ref{eq:TF_spin_prec}). The  $i=1$, 2 subscripts represent the main and minority phase components, respectively. The expected internal-field shift, as calculated by means of equation~(\ref{eq:loc_field}) from the measured dc magnetization $\mu_0 M_{\parallel}$, is reported as well. See text for details.}
\lineup
\begin{footnotesize}
\begin{tabular}{ccccccc} 
\br
$T$ (K) & $B_{\mathrm{ext}}$ (mT) &  $\lambda_1$ ($\mu s^{-1}$) & \textit{a}$_2$ (in wt \%) & $[B_{\mu}-B_{\mathrm{ext}}]_{\mathrm{exp}}$ (mT) & $[B_{\mu}-B_{\mathrm{ext}}]_{\mathrm{calc}}$ (mT) & $\mu_0M_{\parallel}$ (mT) \\
\mr
240 &   50 &  2.56(5) &  2.9(8) &  -4.3(2) & -4.2 & 7.34(1)  \\
 15 &  100 &  5.5(3)  &  2.4(5) &  -7.4(4) & -6.4 & 11.29(3) \\
  5 &  100 &  6.5(2)  &  2.4(4) &  -7.5(3) & -6.1 & 10.77(2) \\
  5 &  125 &  7.6(3)  &  1.8(4) &  -7.7(5) & -6.7 & 11.90(1) \\
\br
\end{tabular}
\end{footnotesize}
\end{table*}
Finally, while keeping the sample at 5\,K, $B_{\mathrm{ext}}$ was raised to 
125\,mT. As expected, the small peak due to the minority normal phase follows 
closely the field change [i.e., it remains unshifted on the relative-field
scale of figure~\ref{fig:normFFT}(b)]. At the same time, the main
peak, related to the SC phase, shows two important features: (\textit{i}) the measured diamagnetic
shift $B_{\mu}-B_{\mathrm{ext}}$ increases due to an increase in bulk magnetization; 
(\textit{ii}) the depolarization rate is further enhanced, as evidenced
by the hatched region on the left side of the main peak in figure~\ref{fig:normFFT}(b).
Yet, the time-domain asymmetry data can still be fully taken into account by 
only two components, one of which with a marginal weight of 2\% 
(see table~\ref{tab:shift}). The presence of the hatched region cannot be exclusively ascribed to the diluted ferromagnetic impurities: if this were the case it would give rise
to a symmetric peak. Evidently, since the experiments show an asymmetric
shape, we conclude that the latter reflects strong pinning effects.

\begin{figure}[tb]
\centering
\includegraphics[width=0.6\textwidth]{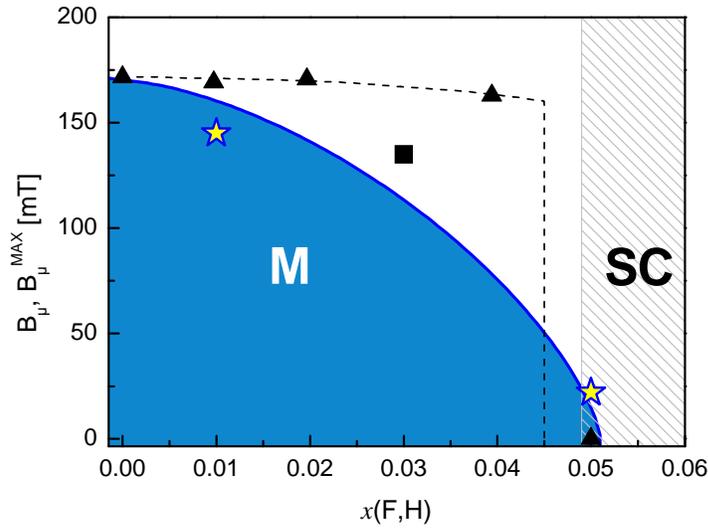}
\caption{\label{fig:phaseDiagram} Magnetic order parameter represented by the maximum value assumed by $B_{\mathrm{\mu}}$ at the lowest temperature as a function of electron doping (either F or H). Triangles represent field values as determined in \cite{Luetkens2009}, the square the value found in \cite{JPCarlo2009}, while the stars refer to the H-doped samples object of the current study. The continuous and dashed lines represent guides for the eyes in the case of a first order and second order transition respectively. The hatched area represents both the onset of bulk superconductivity and the M-SC crossover, where the nanoscopic M-SC  coexistence is realized.}
\end{figure}

In conclusion, the pinning experiment strongly suggests that the superconducting state
\emph{involves almost the whole sample volume}, despite this phase being masked in 
the ZFC susceptibility data [see figure~\ref{fig:SQUID005}(a)]
due to the simultaneous presence of a low-temperature SDW order and of 
diluted ferromagnetic impurities. Since both the SDW order (at 
$T_{\mathrm{N}}< 38$\,K) and the superconducting state (below $T_c \sim 10$\,K)
concern the bulk of the sample, the M-SC coexistence detected below 10\,K is most likely \textit{nanoscopic}, as 
is the case also for Sm-1111 \cite{SannaPRB2009},
Ce-1111 \cite{Sanna2010}, and Nd-1111 \cite{LamuraXXXX}. 
This result hints at a universal picture for the M-SC phase boundary of
Ln-1111 compounds. 
Figure~\ref{fig:phaseDiagram} combines existing litterature data with those arising from the present study.
Since the $x$(F) values for the samples studied in \cite{Luetkens2009} and \cite{JPCarlo2009} 
are nominal, while the $x$(H) values in our case are measured ones, we can only draw a \emph{tentative}
La-1111 phase diagram in the low electron-doping range. Nevertheless, it is important to 
note that because of the $x=0.05$ sample, the magnetic order parameter loses the discontinuity 
at 0.04, thus \emph{pointing to a more probable second-order-like phase transition}.
\section{\label{sec:concl}Conclusion}
Two hydrogen-doped La-1111 polycrystalline samples \cite{Iimura2012} with 
$x(\mathrm{H})=0.01$ and 0.05 were extensively studied by means of dc resistivity, 
dc magnetization and \mbox{ZF-,} TF-, and LF- muon-spin spectroscopy. 
In the $x=0.01$ case our data evidence a long-range SDW order with a N\'eel 
temperature $T_{\mathrm{N}} = 119$\,K. The internal fields detected in this phase are in good agreement with previous results on a $x(\mathrm{F})=0.03$ sample \cite{JPCarlo2009}.
This finding suggests a close analogy between the F-O and H-O substitutions in 
realizing equivalent electron-doping levels in the FeAs planes.

The $x(\mathrm{H})=0.05$ sample too displays an AF magnetic order below
ca.\ 38\,K but, in addition, it shows also bulk superconductivity below $T_c = 10$\,K.
We argue that at low temperature the M and SC phases coexist at a nanoscopic level, 
in contrast with the abrupt M-SC transition previously found in La-1111 compound \cite{Luetkens2009}.
Therefore, it seems that the presence of a \textit{crossover region} is likely a universal 
feature of the electron-doped  Ln-1111 pnictides, which should display \textit{qualitatively 
similar} rare-earth-indepedent phase diagrams, as confirmed, e.g., by 
Sm-1111 \cite{SannaPRB2009}, Ce-1111 \cite{Sanna2010}, and Nd-1111 \cite{LamuraXXXX}.
Yet, the details, such as the extent of the crossover region, could easily depend on 
the rare earth involved.
Further investigations on La-1111 compounds with $x(\mathrm{F})$ between 4 
and 5\% could be interesting to definitely understand the peculiar
role played by lanthanum in determining the M-SC crossover in the Ln-1111 family.

\ack
This work was performed at the Swiss Muon Source S$\mu$S, Paul Scherrer Institut
(PSI, Switzerland) and was in part supported by the Schweizerische Nationalfonds
zur F\"{o}rderung der Wissenschaftlichen Forschung (SNF) and the NCCR research
pool MaNEP of SNF. The authors are grateful to A.~Amato for the instrumental support.
G.L., M.R.C., F.C., and M.P.\ gratefully acknowledge F. Canepa for the careful reading of the manuscript and his fruithful suggestions. This work was partially supported by FP7-EU project SUPER-IRON (No.\ 283204) and by MIUR under project PRIN2012X3YFZ2. S.S.\ acknowledges the financial support from Fondazione Cariplo (research grant no. 2011-0266). The study at Tokyo Tech was supported by JSPS FIRST Program and MEXT Element Strategy Initiative.\\ \\


\begin{thebibliography}{10}
\expandafter\ifx\csname url\endcsname\relax
  \def\url#1{{\tt #1}}\fi
\expandafter\ifx\csname urlprefix\endcsname\relax\def\urlprefix{URL }\fi
\providecommand{\eprint}[2][]{\url{#2}}

\bibitem{Kamihara2008}
Kamihara Y, Watanabe T, Hirano M and Hosono H 2008 {\em J. Am. Chem. Soc.\/}
  {\bf 130}(11) 3296--3297

\bibitem{Luetkens2009}
Luetkens H, Klauss H~H, Kraken M, Litterst F~J, Dellmann T, Klingeler R, Hess
  C, Khasanov R, Amato A, Baines C, Kosmala M, Schumann O~J, Braden M,
  Hamann-Borrero J, Leps N, Kondrat A, Behr G, Werner J and B{\"u}chner B 2009
  {\em Nat. Mater.\/} {\bf 8} 305--309

\bibitem{Kadono2009}
Takeshita S and Kadono R 2009 {\em New J. Phys.\/} {\bf 59} 035006

\bibitem{Khasanov2011}
Khasanov R, Sanna S, Prando G, Shermadini Z, Bendele M, Amato A, Carretta P,
  De~Renzi R, Karpinski J, Katrych S, Luetkens H and Zhigadlo N~D 2011 {\em
  Phys. Rev. B\/} {\bf 84}(10) 100501

\bibitem{Shiroka2011}
Shiroka T, Lamura G, Sanna S, Prando G, {De Renzi} R, Tropeano M, Cimberle M~R,
  Martinelli A, Bernini C, Palenzona A, Fittipaldi R, Vecchione A, Carretta P,
  Siri A~S, Ferdeghini C and Putti M 2011 {\em Phys. Rev. B\/} {\bf 84} 195123

\bibitem{SannaPRB2009}
Sanna S, Renzi R~D, Lamura G, Ferdeghini C, Palenzona A, Putti M, Tropeano M
  and Shiroka T 2009 {\em Phys. Rev. B\/} {\bf 80} 052503

\bibitem{LamuraXXXX}
Lamura G and \textit{et al} 2014 {M}anuscript in preparation

\bibitem{Sanna2011}
Sanna S, Carretta P, Bonf\`a P, Prando G, Allodi G, Renzi R~D, Shiroka T,
  Lamura G, Martinelli A and Putti M 2011 {\em Phys. Rev. Lett.\/} {\bf 107}
  227003

\bibitem{Sanna2013}
Sanna S, Carretta P, Renzi R~D, Prando G, Bonf\`a P, Mazzani M, Lamura G,
  Shiroka T, Kobayashi Y and Sato M 2013 {\em Phys. Rev. B\/} {\bf 87} 134518

\bibitem{Prando2013}
Prando G, Vakaliuk O, Sanna S, Lamura G, Shiroka T, Bonf\`a P, Carretta P,
  Renzi R~D, Klauss H~H, Blum C~G~F, Wurmehl S, Hess C and B{\"u}chner B 2013
  {\em Phys. Rev. B\/} {\bf 87} 174519

\bibitem{Miyazawa2010}
Miyazawa K, Ishida S, Kihou K, Shirage P~M, Nakajima M, Lee C~H, Kito H,
  Tomioka Y, Ito T, Eisaki H, Yamashita H, Mukuda H, Tokiwa K, Uchida S and Iyo
  A 2010 {\em Appl. Phys. Lett.\/} {\bf 96} 072514

\bibitem{Iimura2012}
Iimura S, Matuishi S, Sato H, Hanna T, Muraba Y, Kim S~W, Kim J~E, Takata M and
  Hosono H 2012 {\em Nat. Commun.\/} {\bf 3} 943

\bibitem{kadono2014}
Hiraishi M, Iimura S, Kojima K~M, Yamaura J, Hiraka H, Ikeda K, Miao P,
  Ishikawa Y, Torii S, Miyazaki M, Yamauchi I, Koda A, Ishii K, Yoshida M,
  Mizuki J, Kadono R, Kumai R, Kamiyama T, Otomo T, Murakami Y, Matsuishi S and
  Hosono H 2014 {\em Nat. Physics\/} {\bf 10} 300

\bibitem{JPCarlo2009}
Carlo J~P, Uemura Y~J, Goko T, MacDougall G~J, Rodriguez J~A, Yu W, Luke G~M,
  Dai P, Shannon N, Miyasaka S, Suzuki S, Tajima S, Chen G~F, Hu W~Z, Luo J~L
  and Wang N~L 2009 {\em Phys. Rev. Lett.\/} {\bf 102} 087001

\bibitem{Hanna2011}
Hanna T, Muraba Y, Matsuishi S, Igawa N, Kodama K, Shamoto S and Hosono H 2011
  {\em Phys. Rev. B\/} {\bf 84}(2) 024521

\bibitem{Hess2009}
Hess C, Kondrat A, Narduzzo A, Hamann-Borrero J~E, Klingeler R, Werner J, Behr
  G and B{\"u}chner B 2009 {\em Europhys. Lett.\/} {\bf 87} 17005

\bibitem{Jesche2012}
Jesche A, Nitsche F, Probst S, Doert T, M\"uller P and Ruck M 2012 {\em Phys.
  Rev. B\/} {\bf 86} 134511

\bibitem{Marck2009}
van~der Marck S~C 1997 {\em Phys. Rev. B\/} {\bf 55} 1514

\bibitem{Bernhard2000}
Bernhard C, Tallon J~L, Brucher E and Kremer R~K 2000 {\em Phys. Rev. B\/} {\bf
  61} R14960

\bibitem{Cimberle2003}
Cimberle M~R, Masini R, Ferdeghini C, Artini C and Costa G 2003 {\em Supercond.
  Sci. Technol.\/} {\bf 16} 726

\bibitem{Sanna2010}
Sanna S, De~Renzi R, Shiroka T, Lamura G, Prando G, Carretta P, Putti M,
  Martinelli A, Cimberle M~R, Tropeano M and Palenzona A 2010 {\em Phys. Rev.
  B\/} {\bf 82} 060508(R)

\bibitem{Kohama2009}
Kohama Y, Kamihara Y, Baily A, Civale L, Riggs S~C, Balakirev F~F, Atake T,
  Jaime M, Hirano M and Hosono H 2009 {\em Phys. Rev. B\/} {\bf 79} 144527

\bibitem{Yaouanc2011}
Yaouanc A and {Dalmas de R\'eotier} P 2011 {\em Muon Spin Rotation, Relaxation,
  and Resonance: Applications to Condensed Matter\/} (Oxford: Oxford University
  Press)

\bibitem{Drew2008}
Drew A~J, Pratt F~L, Lancaster T, Blundell S~J, Baker P~J, Liu R~H, Wu G, Chen
  X~H, Watanabe I, Malik V~K, Dubroka A, Kim K~W, Rossle M and Bernhard C 2008
  {\em Phys. Rev. Lett.\/} {\bf 101} 097010

\bibitem{SAVICI}
Savici A~T, Fudamoto Y, Gat I~M, Ito T, Larkin M~I, Uemura Y~J, Luke G~M,
  Kojima K~M, Lee Y~S, Kastner M~A, Birgeneau R~J and Yamada K 2002 {\em Phys.
  Rev. B\/} {\bf 66} 014524 and references therein

\bibitem{Maeter2009}
Maeter H, Luetkens H, Pashkevich Y~G, Kwadrin A, Khasanov R, Amato A, Gusev
  A~A, Lamonova K~V, Chervinskii D~A, Klingeler R, Hess C, Behr G, B{\"u}chner
  B and Klauss H~H 2009 {\em Phys. Rev. B\/} {\bf 80} 094524

\bibitem{Bonfa2012}
Bonf\`a P, Carretta P, Sanna S, Lamura G, Prando G, Martinelli A, Palenzona A,
  Tropeano M, Putti M and {De Renzi} R 2012 {\em Phys. Rev. B\/} {\bf 85}
  054518

\bibitem{Sanna2009}
Sanna S, {De Renzi} R, Lamura G, Ferdeghini C, Martinelli A, Palenzona A, Putti
  M, Tropeano M and Shiroka T 2012 {\em J. Supercond. Nov. Magn.\/} {\bf 22}
  585--588

\bibitem{Bernhard2012}
Bernhard C, Wang C~N, Nuccio L, Schulz L, Zaharko O, Larsen J, Aristizabal C,
  Willis M, Drew A~J, Varma G~D, Wolf T and Niedermayer C 2012 {\em Phys. Rev.
  B\/} {\bf 86} 184509

\bibitem{Pool}
Pool C~H, Farach H~A and Creswick R~J 1995 {\em Superconductivity\/} (London:
  Academic Press)

\bibitem{Cardwell}
Cardwell D~A and Ginley D~S 2003 {\em Handbook of Superconducting Materials\/}
  vol~2 (Bristol: IOP Publishing)

\bibitem{Beleggia2006}
Beleggia M, Graef M~D and Millev Y~T 2006 {\em J. Phys. D: Appl. Phys.\/} {\bf
  39} 891

\bibitem{Luetkens2008PRL}
Luetkens H, Klauss H~H, Khasanov R, Amato A, Klingeler R, Hellmann I, Leps N,
  Kondrat A, Kohler C, Behr G, Werner J and Buchner B 2008 {\em Phys. Rev.
  Lett.\/} {\bf 101} 097009

\end{thebibliography}

\providecommand{\newblock}{}

\end{document}